\documentclass[twocolumn,showpacs,pra,aps,superscriptaddress]{revtex4}
\usepackage{ulem}
\usepackage{array}
\usepackage{graphicx}
\usepackage{epsfig}
\usepackage{amssymb}
\usepackage{amsmath}

\def\bra#1{\langle #1|}
\def\ket#1{|#1 \rangle}
\def\bracket#1#2{\langle #1|#2 \rangle}

\begin{document}
\draft \title{A generalized structure of Bell inequalities for
bipartite
  arbitrary-dimensional systems}

\author{Seung-Woo Lee}

\affiliation{Clarendon Laboratory, University of Oxford, Parks
Road, Oxford OX1 3PU, United Kingdom}

\author{Yong Wook Cheong}

\affiliation{Quantum Photonic Science Research Center, Hanyang
  University, Seoul 133-791, Korea}

\author{Jinhyoung Lee}
\affiliation{Quantum Photonic Science Research Center, Hanyang
  University, Seoul 133-791, Korea}
\affiliation{Department of Physics, Hanyang University, Seoul
133-791, Korea}

\date{\today}

\begin{abstract}

  We propose a generalized structure of Bell inequalities for
  arbitrary $d$-dimensional bipartite systems, which includes the
  existing two types of Bell inequalities introduced by
  Collins-Gisin-Linden-Massar-Popescu [\prl {\bf88}, 040404
  (2002)] and Son-Lee-Kim [\prl {\bf96}, 060406 (2006)]. We
  analyze Bell inequalities in terms of correlation functions and
  joint probabilities, and show that the coefficients of
  correlation functions and those of joint probabilities are in
  Fourier transform relations. We finally show that the
  coefficients in the generalized structure determine the
  characteristics of quantum violation and tightness.

\end{abstract}

%\pacs{03.67.-a,03.67.Dd,03.67.Hk }

\maketitle

\section{Introduction}

Local-realistic theories impose constraints on any correlations
obtained from measurement between two separated systems
\cite{Bell64,CHSH69,tBell}. It was shown that these constraints,
known as Bell inequalities, are incompatible with the quantitative
predictions by quantum mechanics in case of entangled states. For
example, the original Bell inequality is violated by a singlet
state of two spin-1/2 particles \cite{Bell64}. The
Clauser-Horne-Shimony-Holt (CHSH) inequality is another common
form of Bell inequality, allowing more flexibility in local
measurement configurations \cite{CHSH69}. These constraints are of
great importance for understanding the conceptual features of
quantum mechanics and draw the boundary between local-realistic
and quantum correlations. One may doubt if there is any
well-defined constraint for many high-dimensional subsystems which
would eventually simulate a classical system as increasing its
dimensionality to infinity \cite{tBell}. Therefore, constraints
for more complex systems such as multi-partite or high-dimensional
systems have been proposed and investigated intensively
\cite{sWerner,Collins02,Kaszli00,Masanes03,Cerf02,Laskowski04,
Acin02,JLee04,Zohren06,WSon05,WSon06,Brukner02,WSon04,Werner01,Peres99}.

For bipartite high-dimensional systems, Collins {\em et al}.
suggested a local-realistic constraint, called CGLMP inequality
\cite{Collins02}. It is violated by quantum mechanics and its
characteristics of violation are consistent with the numerical
results provided by Kaszlikowski {\em et al.} \cite{Kaszli00}.
Further, Masanes showed that the CGLMP inequality is tight
\cite{Masanes03}, which implies that the inequality has no
interior bias as a local-realistic constraint. However, Acin
\textit{et al.} found that the CGLMP inequality shows maximal
violation by non-maximally entangled state \cite{Acin02}. Zohren and Gill found the similar results when they applied CGLMP inequality to infinite dimensional systems \cite{Zohren06}.
Recently, Son {\em et al.} \cite{WSon05} suggested a generic Bell
inequality and its variant for arbitrary high-dimensional systems.
The variant will be called SLK inequality throughout this paper.
They showed that the SLK inequality is maximally violated by
maximally entangled state. Very recently, the CGLMP inequality was
recasted in the structure of the SLK inequality by choosing
appropriate coefficients \cite{WSon06}.

In this paper, we propose a generalized structure of Bell inequalities
for bipartite arbitrary $d$-dimensional systems, which includes various
types of Bell inequalities proposed previously. A Bell inequality in the
given generalized structure can be represented either in the correlation
function space or joint probability space. We show that a Bell
inequality in one space can be mapped into the other space by Fourier
transformation. The two types of high-dimensional Bell inequalities,
CGLMP and SLK, are represented in terms of the generalized structure
with appropriate coefficients in both spaces (Sec.~\ref{section:GBI}).
We investigate the violation of Bell inequalities by quantum mechanics.
The expectations of local-realistic theories and quantum mechanics are
determined by the coefficients of correlation functions or joint
probabilities.  The CGLMP inequality is maximally violated by
non-maximally entangled state while the SLK is by maximally entangled
state (Sec.~\ref{section:QV}). We also investigate the tightness of Bell
inequalities which represents whether they contain an interior bias or
not at the boundary between local-realistic and quantum correlations.
Then we show that the SLK is a non-tight Bell inequality while the CGLMP
is tight (Sec.~\ref{section:TI}).

\section{Generalized arbitrary dimensional Bell inequality}
\label{section:GBI}

We generalize a Bell inequality for bipartite arbitrary $d$-dimensional
systems. Suppose that each observer independently choose one of two
observables denoted by $A_1$ or $A_2$ for Alice, and $B_1$ or $B_2$ for
Bob. Here we associate a hermitian observables $H$ to a unitary operator
$U$ by the simple correspondence, $U=\exp(iH)$, and call $U$ a unitary
observable \cite{Cerf02,Brukner02,JLee04,WSon05}. We note that unitary
observable representation induces mathematical simplifications without
altering physical results \cite{note}. Each outcome takes the value of an element in
the set of order $d$, $V=\{1,\omega,...,\omega^{d-1}\}$, where
$\omega=\exp(2\pi i/d)$.  The assumption of local-realistic theories
implies that the outcomes of observables are predetermined before
measurements and the role of the measurements is just to reveal the
values. The values are determined only by local hidden variables
$\lambda$, i.e., $A_a(\lambda)$ and $B_b(\lambda)$ for $a,b =1,2$.

We denote a correlation between specific measurements taken by two
observers, as $A_a(\lambda)B^*_b(\lambda)$. Based on the local
hidden-variable description, the correlation function is the average
over many trials of the experiment as
\begin{eqnarray}
  \label{eq:CorrFn1}
  C_{ab} = \int d\lambda ~\rho(\lambda)A_a(\lambda)B^*_b(\lambda),
\end{eqnarray}
where $\rho(\lambda)$ is the statistical distribution of the
hidden variables $\lambda$ with the properties of $\rho(\lambda)
\ge 0$ and $\int d\lambda \rho(\lambda) = 1$. The correlation
function can be expanded in terms of joint probability functions
over all possible outcome pairs $(k,l)$ with complex-valued weight
as
\begin{eqnarray}
  \label{eq:CorrFn2}
  C_{ab} = \sum_{k,l=0}^{d-1}\omega^{k-l}P(A_{a}=k, B_{b}=l),
\end{eqnarray}
where $\omega^{k-l}$ is called a correlation weight and
$P(A_{a}=k, B_{b}=l)$ is a joint probability of Alice and Bob
obtaining outcomes $\omega^k$ and $\omega^l$ respectively. Here we
use the powers $k$ and $l$ of the outcomes $\omega^k$ and
$\omega^l$ for the arguments of the joint probability as there is
one-to-one correspondence.

We assume in general a correlation weight $\mu_{k,l}$ to satisfy
certain conditions \cite{WSon04}. [\textrm{C.1}] The correlation
expectation vanishes for a bipartite system with a locally
unpolarized subsystem: $\sum_{k}\mu_{k,l}=0,\forall l$ and
$\sum_{l}\mu_{k,l}=0,\forall k$ [\textrm{C.2}] The correlation
weight is unbiased over possible outcomes of each subsystem
(translational symmetry within modulo $d$): $
\mu_{k,l}=\mu_{k+\gamma, l+\gamma},\forall \gamma$. [\textrm{C.3}]
The correlation weight is uniformly distributed modulo $d$:
$|\mu_{k+1,l}-\mu_{k,l}|=|\mu_{k,l+1}-\mu_{k,l}|,\forall k,l$. The
correlation weight in Eq.~(\ref{eq:CorrFn2}) $\omega^{k-l}$
satisfies all the conditions, can be written as $\omega^{\alpha}$
where $\alpha\equiv k-l \in\{0,1,...,d-1\}$ and it obeys
$\sum_{\alpha}\omega^{\alpha}=0$.

Let us now consider higher-order($n$) correlations following also
the local hidden-variable description. The $n$-th order
correlation function averaged over many trials of the experiment
corresponds to the $n$-th power of $1$-st order correlation as
\begin{eqnarray}
  \label{eq:CorrFn3}
  C^{(n)}_{ab} &=&  \int d\lambda
  ~\rho(\lambda) \left(A_a(\lambda) B^*_b(\lambda)\right)^n \nonumber \\
  &=&\sum_{k,l=0}^{d-1}\omega^{n(k-l)}P(A_{a}=k, B_{b}=l) \nonumber \\
  &=&\sum_{\alpha=0}^{d-1}\omega^{n\alpha}P(A_{a}\doteq
  B_{b}+\alpha),
\end{eqnarray}
where the $n$-th order correlation weight $\omega^{n\alpha}$ also
satisfies the above conditions, \textrm{C.1}, \textrm{C.2} and
\textrm{C.3}, and $P(A_{a}\doteq B_{b}+\alpha)$ is the joint
probability of local measurement outcomes differing by a positive
residue $\alpha$ modulo $d$. Here we note that the higher-order
correlations Eq.~(\ref{eq:CorrFn3}) show the periodicity of
$C^{(d+n)}_{ab}=C^{(n)}_{ab}$ and they have the Fourier relation
with the joint probabilities as given in Eq.~(\ref{eq:CorrFn3}).

We present a generalized Bell function for arbitrary
$d$-dimensional system using higher-order correlation functions as
\begin{eqnarray}
  \label{eq:GeneralBF}
  {\cal B}=\sum_{a,b}\sum_{n=0}^{d-1}f_{ab}(n)C^{(n)}_{ab},
\end{eqnarray}
where coefficients $f_{ab}(n)$ are functions of the correlation
order $n$ and the measurement configurations $a$, $b$. They
determine the constraint of local-realistic theories with a
certain upper bound and its violation by quantum mechanics will be
investigated in Sec.~\ref{section:QV}. The zero-th order
correlation has no meaning as it simply shift the value of ${\cal
B}$ by a constant and is chosen to vanish, i.e.,
$\sum_{a,b}f_{ab}(0)=0$. The Bell function in
Eq.~(\ref{eq:GeneralBF}) is rewritten in terms of the joint
probabilities given in Eq.~(\ref{eq:CorrFn3}), as
\begin{eqnarray}
  \label{eq:GeneralBF2}
  {\cal B}=\sum_{a,b}\sum_{\alpha=0}^{d-1}\epsilon_{ab}(\alpha)P(A_{a}\doteq B_{b}+\alpha),
\end{eqnarray}
where $\epsilon_{ab}(\alpha)$ are coefficients of the joint
probabilities $P(A_{a}\doteq B_{b}+\alpha)$.

We note that the coefficients $\epsilon_{ab}(\alpha)$ are obtained
by the Fourier transformation of $f_{ab}(n)$ based on the kernel
of a given correlation weight as
\begin{eqnarray}
  \label{eq:CoeffRelation1}
  \epsilon_{ab}(\alpha)&=&\sum_{n=0}^{d-1}f_{ab}(n)\omega^{n\alpha},\\
  \label{eq:CoeffRelation2}
  f_{ab}(n)&=&\frac{1}{d}\sum_{\alpha=0}^{d-1}\epsilon_{ab}(\alpha)\omega^{-n\alpha}.
\end{eqnarray}
It is remarkable that one can represent a given Bell function
either in the correlation function space or joint probability
space by using the Fourier transformation of the coefficients
between them. This is the generalization of the Fourier
transformation in 2-dimensional Bell inequalities provided by
Werner \textit{et al.} \cite{Werner01}.

Different Bell inequalities can be represented by altering
coefficients of the generalized structure, including previously
proposed Bell inequalities in bipartite systems. In the case of
$d=2$, CHSH-type inequalities can be obtained with coefficients as
$f(1)=(1,1,-1,1)$ and
$\epsilon_{ab}(\alpha)=f_{ab}(1)(-1)^{\alpha}$ where
$\alpha\in\{0,1\}$. For arbitrary $d$-dimensional systems, the two
types of Bell inequalities, CGLMP and SLK, are represented in
terms of the generalized structure with appropriate coefficients
obtained as follows.

{\em CGLMP inequality} - As it was originally proposed in terms of
joint probabilities \cite{Collins02}, the Bell function of the
CGLMP inequality is in the form of (\ref{eq:GeneralBF2}) and its
coefficients are given as
\begin{eqnarray}
  \label{eq:CoeffCGLMP}
  \nonumber
  \epsilon_{11}(\alpha)&=&1-\frac{2\alpha}{d-1},~~~
  \epsilon_{12}(\alpha)=-1+\frac{2\dot{(\alpha-1)}}{d-1},\\
  \epsilon_{21}(\alpha)&=&-1+\frac{2\alpha}{d-1},~~~
  \epsilon_{22}(\alpha)=1-\frac{2\alpha}{d-1},
\end{eqnarray}
where the dot implies the positive residue modulo $d$. By using
the inverse Fourier transformation in
Eq.~(\ref{eq:CoeffRelation2}) the coefficients for the correlation
function representation are obtained as
\begin{eqnarray}
  \label{eq:CoeffCGLMP2}
  \nonumber
  f_{11}(n\neq0)&=&\frac{2}{d-1}\left(\frac{1}{1-\omega^{-n}}\right),\\
  \nonumber f_{12}(n\neq0)&=&\frac{2}{d-1}\left(\frac{1}{1-\omega^{n}}\right),\\
  \nonumber f_{21}(n\neq0)&=&\frac{2}{d-1}\left(\frac{1}{\omega^{-n}-1}\right),\\
  \nonumber f_{22}(n\neq0)&=&\frac{2}{d-1}\left(\frac{1}{1-\omega^{-n}}\right),\\
  f_{ab}(n=0)&=&0 ~~~~~\forall a,b,
\end{eqnarray}
where the sum of the 0-th order coefficients vanishes and does not
affect the characteristics of the Bell inequality.

{\em SLK inequality} - It was introduced in terms of correlation
functions \cite{WSon05}, and the coefficients are given by
\begin{eqnarray}
  \label{eq:CoeffSLK}
  \nonumber
  f_{11}(n\neq0)&=&(\omega^{n\delta}+\omega^{(n-d)\delta})/4,\\
  \nonumber f_{12}(n\neq0)&=&(\omega^{n(\delta+\eta_1)}+\omega^{(n-d)(\delta+\eta_1)})/4,\\
  \nonumber f_{21}(n\neq0)&=&(\omega^{n(\delta+\eta_2)}+\omega^{(n-d)(\delta+\eta_2)})/4,\\
  \nonumber f_{22}(n\neq0)&=&(\omega^{n(\delta+\eta_1+\eta_2)}+\omega^{(n-d)(\delta+\eta_1+\eta_2)})/4,\\
  f_{ab}(n=0)&=&0 ~~~~~\forall a,b,
\end{eqnarray}
where $\delta$ is a real number, called a variant factor, and
$\eta_{1,2}\in\{+1/2,-1/2\}$. By varying $\delta$ and
$\eta_{1,2}$, one can have many variants of SLK inequalities. For
all the variants the coefficients in the joint probability picture
are obtained as
\begin{eqnarray}
  \label{eq:CoeffSLK2}
  \nonumber
  \epsilon_{11}(\alpha)&=&S(\delta+\alpha),\\
  \nonumber \epsilon_{12}(\alpha)&=&S(\delta+\eta_1+\alpha),\\
  \nonumber \epsilon_{21}(\alpha)&=&S(\delta+\eta_2+\alpha),\\
  \epsilon_{22}(\alpha)&=&S(\delta+\eta_1+\eta_2+\alpha),
\end{eqnarray}
where
\begin{eqnarray}
  \nonumber
  S(x\neq0)&=&\frac{1}{4}(\cot{\frac{\pi}{d}x}\sin{2\pi x}-\cos{2\pi
  x}-1),\\
  S(x=0)&=&\frac{1}{2}(d-1).
\end{eqnarray}
We have shown that those two types of high-dimensional
inequalities have different coefficients but the same generalized
structure. In the frame work of the generalized structure we will
now study how the coefficients determine the characteristics of
Bell inequalities such as the degree of violation and tightness.

\section{Violation by Quantum Mechanics}
\label{section:QV}

In order to see the violation of Bell inequalities by quantum
mechanics we need to know the upper bound by local hidden variable
theories. We note that a probabilistic expectation of a Bell
function is given by the convex combination of all possible
deterministic values of the Bell function and the local-realistic
upper bound is decided by the maximal deterministic value. Let
$\alpha_{ab} = \alpha$ such that $P(A_a \doteq B_b + \alpha) = 1$.
The assumption of local-realistic theories implies that the values
$\alpha_{ab}$ are predetermined. For a Bell function in the form
of Eq.~(\ref{eq:GeneralBF}) or (\ref{eq:GeneralBF2}), they obey
the constraint, $\alpha_{11}+\alpha_{22}\doteq
\alpha_{12}+\alpha_{21}$, because of the identity, $A_1 - B_1 +
A_2 -B_2 = A_1 - B_2 + A_2 - B_1$. The local-realistic upper bound
of the Bell function is therefore given by

\begin{eqnarray}
  \label{eq:LRmax}
{\cal
B}^{\mathrm{max}}_{\mathrm{LR}}=\max_{\alpha_{ab}}[\sum_{a,b}\epsilon_{ab}(\alpha_{ab})|\alpha_{11}+\alpha_{22}\doteq
\alpha_{12}+\alpha_{21}].
\end{eqnarray}

The quantum expectation value for arbitrary quantum state
$\hat{\rho}$ is written by
\begin{eqnarray}
  \label{eq:QBvalue}
  \nonumber
  {\cal B}_{\mathrm{QM}}(\hat{\rho})&=&\mathrm{Tr}(\hat{{\cal B}}\hat{\rho})\\
  &=&\sum_{a,b}\sum_{n=0}^{d-1}f_{ab}(n)\mathrm{Tr}(\hat{C}_{ab}^{(n)}\hat{\rho}),
\end{eqnarray}
where $\hat{{\cal B}}$ is the Bell operator defined by replacing
the correlation function in Eq.~(\ref{eq:GeneralBF}) with
correlation operator,
$\hat{C}_{ab}^{(n)}=\sum_{k,l}\omega^{n(k-l)}\hat{P}_a\otimes\hat{P}_b$
where $\hat{P}_a$,$\hat{P}_b$ are projectors onto the measurement
basis denoted by $a,b$. If an expectation value of any quantum
state exceeds the local realistic bound ${\cal
B}_{\mathrm{LR}}^{\mathrm{max}}$, i.e., the Bell inequality is
violated by quantum mechanics, the composite system is entangled
and shows nonlocal quantum correlations. The maximal quantum
expectation is called quantum maximum ${\cal
B}_{\mathrm{QM}}^{\mathrm{max}}$ and corresponds to the maximal
eigenvalue of the Bell operator. In the case of $d=2$, with the
coefficients $f(1)=(1,1,-1,1)$ we can obtain the quantum maximum,
${\cal B}_{\mathrm{QM}}^{\mathrm{max}}=2\sqrt{2}$, which is in
agreement with the Cirel'son bound \cite{Cirel80}.

In the presence of white noise, a maximally entangled state
$|\psi_m \rangle$ becomes $\hat{\rho}=p|\psi_{m} \rangle \langle
\psi_{m}|+(1-p)\openone/d^2$ where $p$ is the probability that the
state is unaffected by noise. Then, the minimal probability for
the violation is $p^{\mathrm{min}}={\cal
B}^{\mathrm{max}}_{\mathrm{LR}}/{\cal B}_{\mathrm{QM}}(|\psi_{m}
\rangle)$. We now investigate the violation of two types of Bell
inequalities, CGLMP and SLK, and compare them as follows.

\begin{figure}
\begin{center}
\includegraphics[width=0.45\textwidth]{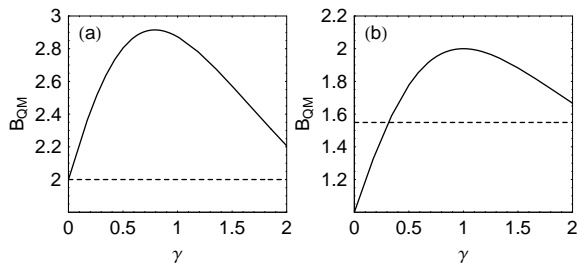}
\caption{The expectation value of (a) the CGLMP and (b) the
optimal SLK for $d=3$ as varying the value $\gamma$ for the
quantum state,
$(1/\sqrt{n})(|00\rangle+\gamma|11\rangle+|22\rangle)$ where
$n=2+\gamma^2$. The SLK takes the maximum $2$ when the state is
maximally entangled ($\gamma=1$), whereas the CGLMP takes the
maximum $2.9149$ for a partially entangled state
($\gamma\simeq0.7923$). The dashed lines indicate the
local-realistic upper bounds.} \label{fig1}
\end{center}
\end{figure}

{\em CGLMP inequality} - The local-realistic upper bound, ${\cal
  B}^{\mathrm{max}}_{\mathrm{LR}}=2$, can be obtained as
Eq.~(\ref{eq:LRmax}). The quantum expectation can also be obtained as
Eq.~(\ref{eq:QBvalue}) and it is consistent with the result in
Ref.~\cite{Collins02}. Acin \textit{et al.} found, however, that the
CGLMP inequality shows maximal violation for non-maximally entangled
states \cite{Acin02}. For $3$-dimensional system, the quantum maximum is
${\cal B}_{\mathrm{QM}}^{\mathrm{max}}\simeq2.9149$ for the
non-maximally entangled state,
$(1/\sqrt{n})(|00\rangle+\gamma|11\rangle+|22\rangle)$ where
$\gamma\simeq0.7923$ and $n=2+\gamma^2$. It is higher than the
expectation by maximally entangled state, ${\cal B}(|\psi_{m}
\rangle)\simeq2.8729$. The expectation of the CGLMP is shown in
Fig.~\ref{fig1} against the entanglement degree $\gamma$, once the local
measurements are chosen such that they maximize the Bell function for
the maximally entangled state. Further, we also note that the minimal
violation probability($p^{\mathrm{min}}$) of the CGLMP decreases as the
dimension $d$ increases.

{\em SLK inequality} - Many variants of the SLK Bell inequality are
obtained by varying $\delta$ and $\eta_{1,2}$. All the variants of the
SLK have the same quantum maximum $d-1$ for a maximally entangled state
$|\psi_m\rangle$, ${\cal B}^{\mathrm{max}}_{\mathrm{QM}}={\cal
  B}_{\mathrm{QM}}(|\psi_{m} \rangle)$, as we prove in Appendix
\ref{app:QUB}. On the other hand, the local-realistic upper bounds
depend on the variants. The local-realistic upper bound ${\cal
  B}^{\mathrm{max}}_{\mathrm{LR}}$ is a function of the variant factor
$\delta$. It shows a periodicity, ${\cal
  B}^{\mathrm{max}}_{\mathrm{LR}}(\delta)={\cal
  B}^{\mathrm{max}}_{\mathrm{LR}}(\delta+1/2)$, and without loss of
generality it suffices to consider $0\leq\delta < 1/2$. If $\delta=0$,
the local-realistic upper bound is the same as the quantum maximum
$d-1$, and thus the corresponding Bell inequality is not violated by
quantum mechanics. When $\delta=1/4$, we have the lowest local-realistic
upper bound as
\begin{eqnarray}
\label{eq:lowestLRUB} \min_{\delta}[{\cal
B}^{\mathrm{max}}_{\mathrm{LR}}(\delta)]=
\frac{1}{4}(3\cot{\frac{\pi}{4d}}-\cot{\frac{3\pi}{4d}})-1,
\end{eqnarray}
and for other cases the bound values are symmetric at $\delta=1/4$,
i.e., ${\cal B}^{\mathrm{max}}_{\mathrm{LR}}(1/4+\epsilon)={\cal
  B}^{\mathrm{max}}_{\mathrm{LR}}(1/4-\epsilon)$ for
$0<\epsilon\leq1/4$. Therefore, we will call the variant of
$\delta=1/4$, which gives the maximal difference between quantum maximum
and local-realistic upper bound, as the {\em optimal} SLK inequality and
use it for comparing to the CGLMP. In Fig.~\ref{fig1}, we present the
quantum expectation of the SLK for 3-dimensional systems against the
degree $\gamma$, where the local measurements are chosen such that they
maximize the Bell function for the maximally entangled state. Note that
the SLK inequality shows the maximal violation by maximally entangled
states and the minimal violation probability $p^{\mathrm{min}}$
increases as the dimension $d$ increases.

By investigating the violation of two inequalities, CGLMP and SLK,
based on the generalized structure of Bell inequalities, we showed
that those two types have very different characteristics. The SLK
inequality is maximally violated by maximally entangled states as
being consistent with our intuition whereas the CGLMP is maximally
violated by non-maximally entangled states. We remark that the
coefficients of the given generalized structure determine the
characteristics of quantum violations.

\section{Tightness of Bell inequalities}
\label{section:TI}

The set of possible outcomes for a given measurement setting forms
a convex polytope in the joint probability space or alternatively
in the correlation function space
\cite{Peres99,Werner01,Masanes03,Laskowski04}. Each generator of
the polytope represents the predetermined measurement outcome
called local-realistic configuration. All interior points of the
polytope are given by the convex combination of generators and
they represent the accessible region of local-realistic theories
associated with the probabilistic expectations of measurement
outcomes. Therefore, every facet of the polytope is a boundary of
halfspace characterized by a linear inequality, which we call
\textit{tight Bell inequality}. There are non-tight Bell
inequalities which contain the polytope in its halfspace. As the
non-tight Bell inequality has interior bias at the boundary
between local-realistic and quantum correlations, one might say it
to be the worse detector of the nonlocal test
\cite{Werner01,Masanes03,Laskowski04}.

The Bell polytope is lying in the joint probability space of
dimension $h$, the degrees of freedom for the measurement raw
data. For a bipartite system, two observables per party and
$d$-dimensional outcomes, the joint probability, $P(A_a=k,B_b=l)$
where $k,l=0,1,...,d-1$ and $a,b=1,2$, can be arranged in a
$4d^2$-dimensional vector space. However, the joint probabilities
have two constraints, i.e., normalization and no-signaling
constraints, which reduce dimension by $4d$ \cite{Masanes03}. The
generators in the $h$-dimensional space can be written, following
the notations in Ref.~\cite{Masanes03}, as
\begin{equation}
\label{eq:generator}
\mathbf{G}=|A_1,B_1\rangle\oplus|A_1,B_2\rangle\
\oplus|A_2,B_1\rangle\ \oplus|A_2,B_2\rangle\,
\end{equation}
where $|n\rangle$ stands for $|n \mod d\rangle$ and is the
$d$-dimensional vector with a 1 in the $n$-th component and $0$s
in the rest.

In order to examine the tightness of a given generalized Bell
inequality, in general one considers the following conditions that
every tight Bell inequality fulfills \cite{Masanes03}.
[\textrm{T.1}] All the generators must belong to the half space of
a given facet. [\textrm{T.2}] Among the generators on the facet,
there must be $h$ which are linearly independent. First, it is
straightforward that all generators fulfill the inequality as the
Bell inequality derived to do. As the local-realistic upper bound
is the maximum among expectation values of local-realistic
configurations, all generators are located below the
local-realistic upper bound, ${\cal
B}^{\mathrm{max}}_{\mathrm{LR}}$. Thus the first condition
\textrm{T.1} is fulfilled. Second, we examine whether there are
$h$ linearly independent generators which give the value of the
local-realistic bound, ${\cal B}^{\mathrm{max}}_{\mathrm{LR}}$. By
the predetermined local-realistic values $\alpha_{ab}$, the
generators (\ref{eq:generator}) become
\begin{eqnarray}
\nonumber &|A,A-\alpha_{11}\rangle\oplus|A,A-\alpha_{12}\rangle
\oplus|A-\alpha_{12}+\alpha_{22},A-\alpha_{11}\rangle\ \\
&\oplus|A-\alpha_{11}+\alpha_{21},A-\alpha_{12}\rangle,
\end{eqnarray}
where $A\in\{0,1,...,d-1\}$ and the number of linearly independent
generators is determined by the number of sets $\{\alpha_{ab}\}$
that give the local-realistic upper bound. If the number of linear
independent generators is not smaller than $h=4d(d-1)$, the
corresponding Bell inequality is tight.

{\em CGLMP inequality} - For the CGLMP inequality the
local-realistic upper bound is achieved when
$\alpha_{11}+\alpha_{22}-\dot{(\alpha_{12}-1)}-\alpha_{21}+d-1=0$.
The condition allows the sufficient number of linearly independent
generators and the CGLMP inequality is tight \cite{Masanes03}.

{\em SLK inequality} - For the optimal SLK inequality, the upper
bound is obtained in the case that
$\{\alpha_{11},\alpha_{12},\alpha_{21},\alpha_{22}\}$ is equal to
one of four sets; $\{0,0,d-1,d-1\}$,
$\{0,0,0,0\}$,$\{0,1,d-1,0\}$,$\{d-1,0,d-1,0\}$. Thus there are
four types of generators as
\begin{eqnarray}
\nonumber &&|A,A\rangle\oplus|A,A\rangle\ \oplus|A-1,A\rangle\
\oplus|A-1,A\rangle\,\\
\nonumber &&|A,A\rangle\oplus|A,A\rangle\ \oplus|A,A\rangle\
\oplus|A,A\rangle\,\\
\nonumber &&|A,A\rangle\oplus|A,A-1\rangle\ \oplus|A-1,A\rangle\
\oplus|A-1,A-1\rangle\,\\
&&|A,A+1\rangle\oplus|A,A\rangle\ \oplus|A,A+1\rangle\
\oplus|A,A\rangle\,
\end{eqnarray}
which are linearly independent with $A\in\{0,1,...,d-1\}$. There
are only $4d$ linearly independent generators which are smaller
than $h=4d(d-1)$, the tightness condition T.2. Thus the optimal
SLK inequality is non-tight. On the other hand, the SLK inequality
for $\delta=0$ is tight but it is not violated by quantum
mechanics.

\section{Remarks}

In summary, we presented a generalized structure of the Bell
inequalities for arbitrary $d$-dimensional bipartite systems by
considering the correlation function specified by a well-defined
complex-valued correlation weight. The coefficients of a given
Bell inequality in the correlation function space and the joint
probability space were shown to be in the Fourier relation. Two
known types of high-dimensional Bell inequalities, CGLMP and SLK,
were shown to have the generalized structure in common and we
found their coefficients in both spaces.

Based on the generalized structure, we investigated
characteristics of the Bell inequalities such as quantum violation
and tightness. We found that the CGLMP and SLK inequalities show
different characteristics. For instance, the SLK inequality is
maximally violated by maximally entangled states, which is
consistent with the intuition ``the larger entanglement, the
stronger violation against local-realistic theories,'' whereas the
CGLMP inequality is maximally violated by the non-maximally
entangled state as previously shown by Acin {\em et
  al.} \cite{Acin02}. On the other hand, in analyzing the tightness of
the inequalities, the CGLMP is tight but the SLK inequality is
found to be non-tight for $\delta\neq 0$, implying that the SLK
inequality has interior bias at the boundary between
local-realistic and quantum correlations.

The correlation coefficients of Bell inequalities play a crucial role in determining their characteristics of quantum violation and tightness. This implies that by altering the coefficients in the generalized structure one can construct other Bell inequalities. The present work opens a possibility of finding a new Bell inequality that fulfills both conditions of the maximal violation by maximal entanglement and the tightness.

\acknowledgments

The authors thanks D. Jaksch for useful discussions. This work was supported by
MOST/KOSEF through the Quantum Photonic Science Research Center and the Korean Research
Foundation Grant funded by the Korean Government (MOEHRD) (KRF-2005-041-C00197).
S.-W.Lee was supported by the EU through the STREP project OLAQUI.

\appendix*

\section{Quantum Maximum of all variant SLK inequalities}
\label{app:QUB}

We shall prove that all the variant SLK inequalities take $d-1$ as
the quantum maximum. The Bell operator of the SLK variants can be
written as
\begin{eqnarray}
\hat{{\cal
B}}_S&=&\frac{1}{2}\sum^{d-1}_{n=1}\boldsymbol{\alpha}\cdot\boldsymbol{\tilde\beta},
\end{eqnarray}
where $\boldsymbol{\alpha}=(\hat{A}^{\dagger n}
_1,\hat{A}^{\dagger
  n}_2)^{T}$ and $\boldsymbol{\tilde\beta}=\mathbf{U}\boldsymbol{\beta}$
with $\boldsymbol{\beta}=(\hat{B}^n_1,\hat{B}^n_2)^{T}$ and
$\mathbf{U}$ is a $2 \times 2$ unitary matrix with elements,
\begin{eqnarray}
\nonumber U_{11}&=&(\omega^{n\delta}+\omega^{(n-d)\delta})/2,\\
\nonumber U_{12}&=&(\omega^{n(\delta+\eta_1)}+\omega^{(n-d)(\delta+\eta_1)})/2,\\
\nonumber U_{21}&=&(\omega^{n(\delta+\eta_2)}+\omega^{(n-d)(\delta+\eta_2)})/2,\\
U_{22}&=&(\omega^{n(\delta+\eta_1+\eta_2)}+\omega^{(n-d)(\delta+\eta_1+\eta_2)})/2,
\end{eqnarray}
where $\eta_{1,2}\in\{1/2,-1/2\}$.

The expectation of the Bell operator is given by
\begin{eqnarray}
  \label{eq:qefvbo}
  &&\frac{1}{2}\left|\sum_{n=1}^{d-1} \bra{\psi} \boldsymbol{\alpha} \cdot
    \boldsymbol{\tilde\beta} \ket{\psi} \right| \le
  \frac{1}{2}\sum_{n=1}^{d-1} \left| \bra{\psi} \boldsymbol{\alpha} \cdot
    \boldsymbol{\tilde\beta} \ket{\psi} \right| \nonumber \\
  &\le&
  \frac{1}{2}\sum_{n=1}^{d-1} \left(\left| \bra{\psi} \alpha_1 \otimes {\tilde\beta}_1
  \ket{\psi} \right| + \left| \bra{\psi} \alpha_2 \otimes {\tilde\beta}_2
  \ket{\psi}\right| \right) \nonumber \\
  &\le& \frac{1}{\sqrt{2}}
  \sum_{n=1}^{d-1} \sqrt{\left| \bra{\psi} \alpha_1 \otimes {\tilde\beta}_1
  \ket{\psi} \right|^2 + \left| \bra{\psi} \alpha_2 \otimes {\tilde\beta}_2
  \ket{\psi}\right|^2} \nonumber \\
  &=&\frac{1}{\sqrt{2}}
  \sum_{n=1}^{d-1} \sqrt{\sum_{i=1}^2 \left| \bra{\psi} \alpha_i \otimes
  {\tilde\beta}_i  \ket{\psi} \right|^2},
\end{eqnarray}
where we consecutively used the triangle inequality and the
arithmetic-geometric means inequality, $ 2 |a||b| \le
|a|^2+|b|^2$. Note that
\begin{eqnarray}
  \nonumber
  \sum_{i=1}^2 \left| \bra{\psi} \alpha_i \otimes {\tilde\beta}_i
  \ket{\psi} \right|^2 &\le& \sum_{i=1}^2 \bra{\psi} \left(\alpha_i^\dag
  \otimes {\tilde\beta}_i^\dag \right) \left(\alpha_i \otimes {\tilde\beta}_i
  \right) \ket{\psi}\\
  &=&\sum_{i=1}^2 \bra{\psi} \openone \otimes {
  \tilde\beta}_i^\dag {\tilde\beta}_i \ket{\psi},
\end{eqnarray}
where we used that $\alpha_i$ is unitary. Here the above
inequality is obtained by reasoning that $\hat{Q} \equiv \openone
- \ket{\psi}\bra{\psi}$ is a positive operator as $\bra{\phi}
\hat{Q} \ket{\phi} = 1 - |\bracket{\phi}{\psi}|^2 \ge 0$ for any
$\ket{\phi}$, and $|\bra{\psi} \hat{C} \ket{\psi}|^2=\bra{\psi}
\hat{C}^\dag \ket{\psi}\bra{\psi} \hat{C} \ket{\psi} = \bra{\psi}
\hat{C}^\dag (\openone - \hat{Q}) \hat{C} \ket{\psi} =
\bra{\psi}\hat{C}^\dag\hat{C} \ket{\psi} -
\bra{\psi}\hat{Q}\ket{\psi} \le \bra{\psi}\hat{C}^\dag\hat{C}
\ket{\psi}$, where $\hat{C}\equiv \alpha_i \otimes
{\tilde\beta}_i$. Since $\sum_i \tilde{\beta}_i^\dag
\tilde{\beta}_i = \sum_{jk} \sum_{i} U_{ij}^* U_{ik} \beta_j^\dag
\beta_k = \sum_{jk} \delta_{jk} \beta_j^\dag \beta_k = \sum_{i}
\beta_i^\dag \beta_i = 2 \openone$, it is clear that
\begin{eqnarray}
  \sum_{i=1}^2 \left| \bra{\psi} \alpha_i \otimes \tilde{\beta}_i
  \ket{\psi} \right|^2 \le \bra{\psi} \openone \otimes \sum_{i=1}^2 {
  \beta}_i^\dag {\beta}_i \ket{\psi}= 2.
\end{eqnarray}
Hence the upper bound for all variants of the SLK is
\begin{eqnarray}
\left|\bra{\psi} \hat{{\cal B}}_S \ket{\psi} \right| \le d-1.
\end{eqnarray}
Since all SLK Bell operators have the eigenvalue $d-1$ for
maximally entangled states, the upper bound is reachable.
Therefore, $d-1$ is the quantum maximum for all variants of the
SLK inequality.


\begin{thebibliography}{99}
\bibitem{note} Instead of the complex eigenvalues, one may consider real eigenvalues but then he/she has to employ rather complicated form of correlation weights to obtain Bell inequalities equivalent to ones derived in the present paper.
\end{thebibliography}
\end{document}